\documentclass{article}

\usepackage{arxiv}

\usepackage[utf8]{inputenc} 
\usepackage[T1]{fontenc}    
\usepackage{hyperref}       
\usepackage{url}            
\usepackage{booktabs}       
\usepackage{amsfonts}       
\usepackage{nicefrac}       
\usepackage{microtype}     
\usepackage{graphicx}
\usepackage{doi}

\usepackage[style=numeric,sorting=none]{biblatex} 
\title{Binamix - A Python Library for Generating Binaural Audio Datasets}

\author{Dan Barry\\
	School of Computer Science\\
	University College Dublin\\
	Dublin, Ireland\\
	\texttt{danbarry@duck.com} \\	
    \And
	Davoud Shariat Panah\\
	School of Computer Science\\
	University College Dublin\\
	Dublin, Ireland\\
	\texttt{davoud.shariatpanah@ucd.ie} \\
    \And
	Alessandro Ragano\\
	School of Computer Science\\
	University College Dublin\\
	Dublin, Ireland\\
	\texttt{alessandro.ragano@ucd.ie} \\
    \And
	Jan Skoglund\\
	Google LLC\\
        San Francisco, CA\\
	\texttt{jks@google.com} \\
    \And
	Andrew Hines\\
	School of Computer Science\\
	University College Dublin\\
	Dublin, Ireland\\
	\texttt{andrew.hines@ucd.ie} \\
}

\renewcommand{\shorttitle}{Binamix}

\hypersetup{
}

\begin{document}
\maketitle

\begin{abstract}
The increasing demand for spatial audio in applications such as virtual reality, immersive media, and spatial audio research necessitates robust solutions to generate binaural audio data sets for use in testing and validation. Binamix is an open-source Python library designed to facilitate programmatic binaural mixing using the extensive SADIE II Database, which provides Head Related Impulse Response (HRIR) and Binaural Room Impulse Response (BRIR) data for 20 subjects. The Binamix library provides a flexible and repeatable framework for creating large-scale spatial audio datasets, making it an invaluable resource for codec evaluation, audio quality metric development, and machine learning model training. A range of pre-built example scripts, utility functions, and visualization plots further streamline the process of custom pipeline creation. This paper presents an overview of the library's capabilities, including binaural rendering, impulse response interpolation, and multi-track mixing for various speaker layouts. The tools utilize a modified Delaunay triangulation technique to achieve accurate HRIR/BRIR interpolation where desired angles are not present in the data. By supporting a wide range of parameters such as azimuth, elevation, subject Impulse Responses (IRs), speaker layouts, mixing controls, and more, the library enables researchers to create large binaural datasets for any downstream purpose. Binamix empowers researchers and developers to advance spatial audio applications with reproducible methodologies by offering an open-source solution for binaural rendering and dataset generation. We release
the library under the Apache 2.0 License at \texttt{\href{https://github.com/QxLabIreland/Binamix/}{https://github.com/QxLabIreland/Binamix/}}

\end{abstract}

\renewcommand\thefootnote{}\footnotetext{Accepted to the 158th Audio Engineering Society Convention}
\addtocounter{footnote}{-1}


\section{Introduction}
Spatial audio has seen a significant rise in popularity across various domains in recent years, from virtual reality and immersive media to auditory research and perceptual evaluation. The release of new audio formats such as Dolby Atmos \cite{DolbyAtmos} and Google Eclipsa \cite{iamf} provide the necessary platforms on which the modern era of spatial audio is being built. A significant rise in media consumption through headphones over the last decade \cite{rane2022survey} is also contributing to advances in the binaural rendering technology required to deliver spatial audio over headphones. As research in the areas of spatial audio codec creation and objective quality metrics \cite{ambiqual} grows, the need for large-scale spatial audio data creation also increases. Such data is essential for testing and validation in codec and quality metric design. Additionally, large spatial audio datasets can be used to train machine learning algorithms for tasks like machine listening and no-reference quality estimation. Although there are existing tools for interfacing with HRIR libraries \cite{pysofaconventions} and binaural rendering \cite{franck2018binaural}, there is no existing solution which combines them together with robust binaural mixing tools, surround speaker layout emulation, and dataset generation.

\section{Binamix}
\label{sec:binamix}
Here, we present Binamix \cite{Binamix2025}, an open-source Python 3 library for binaural audio dataset generation. The library contains a set of pipeline tools for programmatic binaural mixing, leveraging the extensive SADIE II Database [7] which contains high-resolution Head Related Impulse Response (HRIR) and Binaural Room Impulse Response (BRIR) data for 20 subjects. SADIE II contains two dummy head subjects with 8802 IR sample points around the sphere and 18 human subjects with up to 2818 IR sample points around the sphere for the HRIRs and a more sparse sphere sampling for the BRIRs. SADIE was chosen due to its widespread use in both research and commercial applications but Binamix can be used with other HRIR datasets by providing your own scripts to interface with the IR files.

\subsection{Motivation}
\label{subsec:motivation}
Our main work is principally focused on building audio quality metrics \cite{ambiqual}\cite{Hines2015ViSQOL}\cite{NOMAD}\cite{ragano2024scoreq} and as such we require large amounts of labeled data for testing and training. Datasets do exist for testing traditional audio quality metrics \cite{TCD-VOIP} but no such binaural audio dataset exists for testing spatial quality metrics. It is impractical to create enough data manually using spatial audio production tools in a Digital Audio Workstation (DAW), and furthermore, generating all of the test condition variations (subjects, speaker layouts, mixing variations, source locations, etc.) is demanding and requires significant domain knowledge. We developed the Binamix library to generate large amounts of binaural data for our work, but we believe it has many additional applications in spatial audio research. We decided to create our own direct binaural rendering pipeline instead of using an existing Ambisonics renderer. The rationale is that we need to measure the sensitivity of our algorithms to a change in a single variable (azimuth, elevation, subject, speaker layout etc.) without having to account for any spatial distortions arising from the Ambisonics encode/decode process.

\subsection{Tools Overview}
The tools include functions for binaurally mixing and rendering sources for any given azimuth and elevation, even if the SADIE II subject data does not have discrete impulse responses for the desired angles. In this instance, the tools will automatically generate suitable binaural impulse responses at render time by identifying the nearest available angles using Delaunay triangulation \cite{Delaunay} and then using Euclidean distance measures to calculate the interpolation weights. Various interpolation modes can be selected, depending on user requirements, which will be described in more detail in section \ref{fig:interpolation}. This approach ensures that the rendered audio closely matches the intended spatial positioning.

The tools can also simulate binaurally rendered channel-encoded surround sound by simulating vector-based amplitude panning methods (VBAP)\cite{Pulkki1997:VBAPbase} within our binaural mixing pipeline. This simulates the rendering of sources at virtual speaker positions between the discrete speaker locations defined in a surround speaker layout. As such, Binamix allows for simulating channel-encoded surround mixes which have been binaurally rendered. This was motivated by the fact that much of the legacy content which is still in use today, exists in channel-encoded surround formats such as 5.1 and 7.1. These legacy formats will ultimately become the source format for direct binaural rendering when delivered over headphones. Because of this, it is necessary to be able to simulate a wide variety of these conditions when building codecs and quality metrics for binaural audio. The tools also support binaural rendering of real-world channel-encoded surround mixes, if you have access to them. 

The library also includes a wide range of useful SADIE II wrapper functions which can be utilized to create custom pipelines and a range of example scripts to get started. Binamix also includes a loader script for the DSD100 Stems Database \cite{SiSEC16}, which contains individual stems for vocals, drums, bass, and other instruments. However, the tools can be used with any audio content. This paper introduces the main concepts of Binamix and serves as a starting point for using the library. The next section illustrates some usage examples of Binamix.

\subsection{Applications}
In general, the library allows you to achieve almost any task requiring programmatic and repeatable binaural mixing or rendering. Specific use cases include:

\begin{itemize}
  \item Generating binaural and stereo mixes for general testing and evaluation.
  \item Simulating various binaural reference and test conditions for audio codec creation and evaluation.
  \item Simulating various binaural reference and test conditions for audio quality metric creation and evaluation.
  \item Generating large amounts of real-world binaural source renders or binaural mix renders for training machine learning models.
  \item Creating training datasets with a variety of conditions for:
  \begin{itemize}
    \item Azimuth
    \item Elevation
    \item Subject ID (e.g., SADIE subjects; D1, D2, H3... H20)
    \item IR type (HRIR or BRIR)
    \item Speaker layout (e.g., none, 5.1, 7.1, 7.1.4, custom)
    \item Mix Parameters (level, reverb, etc.)
    \item Sample Rate (e.g., 44100, 48000, 96000)
    \item Audio content (any audio you want to mix or render)
  \end{itemize}
\end{itemize}

By offering access to repeatable, programmatic binaural mixing and rendering, the Binamix library supports the development of new technologies and applications in spatial audio.

\section{Method and Design}

\begin{figure}
  \centering
  \includegraphics[width=0.8\textwidth]{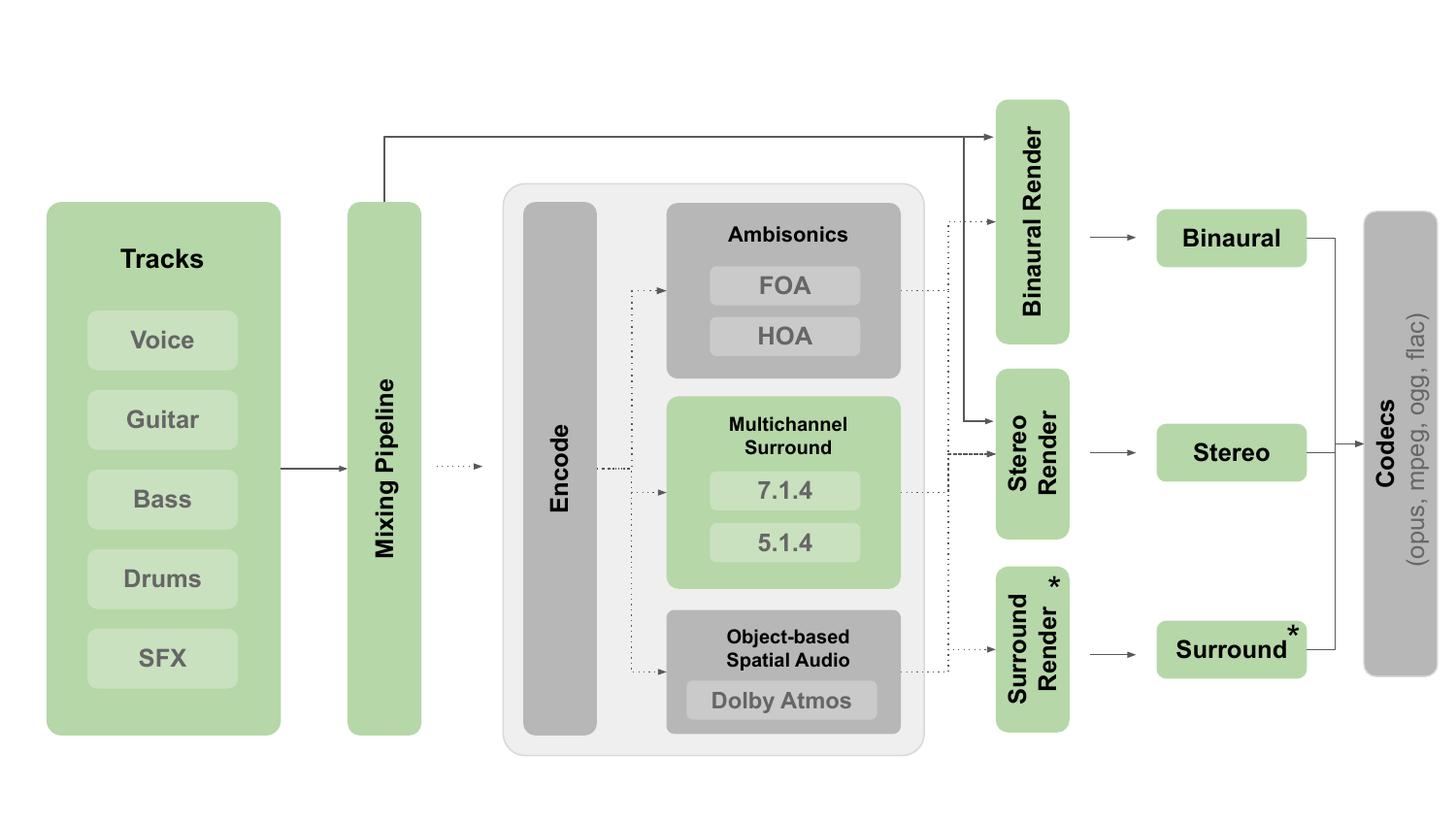}
  \caption{Binamix system overview. Blocks in green are currently supported or *coming soon.}
  \label{fig:overview}
\end{figure}

Binamix is designed to emulate many of the basic functions in a commercial spatial audio processing pipeline. Figure \ref{fig:overview} shows a simplified processing pipeline overview with Binamix components depicted in green. The components we do not currently provide are depicted in gray for reasons discussed in section \ref{subsec:motivation}. 

\subsection{Mixing Architecture}
\begin{figure}
  \centering
  \includegraphics[width=0.8\textwidth]{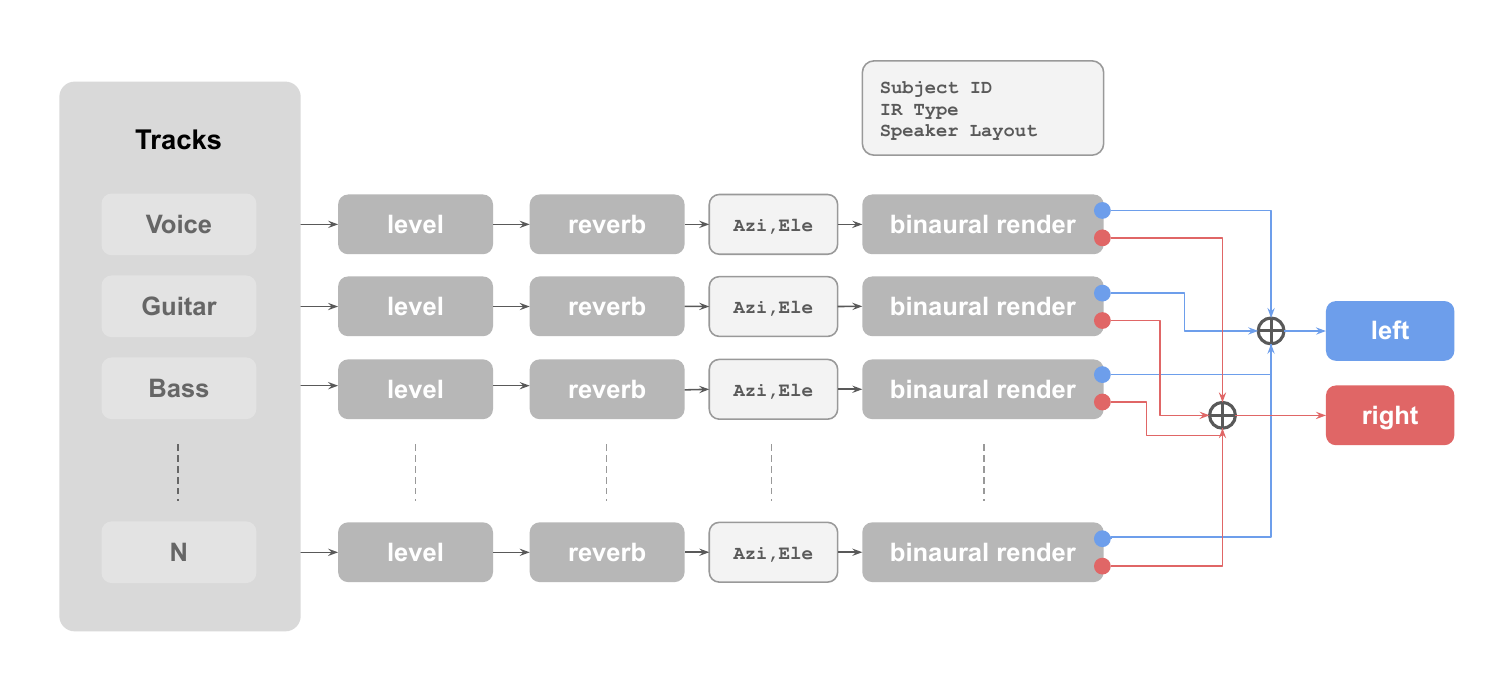}
  \caption{Binamix signal flow.}
  \label{fig:mixer}
\end{figure}

The mixing architecture, shown in Figure \ref{fig:mixer}, allows you to render single sources or binaurally mix multiple sources. When mixing multiple sources, a simple \texttt{TrackObject} class allows you to specify track name, level, reverb, azimuth, and elevation. Level and reverb are specified in the normalized range 0 to 1. Azimuth can be specified either in the range of 0 to 360 degrees or as positive and negative angles. Positive angles are interpreted in the counter-clockwise direction and negative angles are interpreted in the clockwise direction. Elevation is best specified in the range -90 to +90 degrees but the library will convert values outside that range. If the SADIE II Database or selected speaker layout does not contain the discrete angle you specify, Binamix will automatically generate a suitable IR using the interpolation methods described in section \ref{subsec:interpolation}. You can specify as many tracks as you like in this manner. The tracks are then collected into an array of tracks. 

\begin{figure}[ht]
  \centering
    \includegraphics[width=1\textwidth]{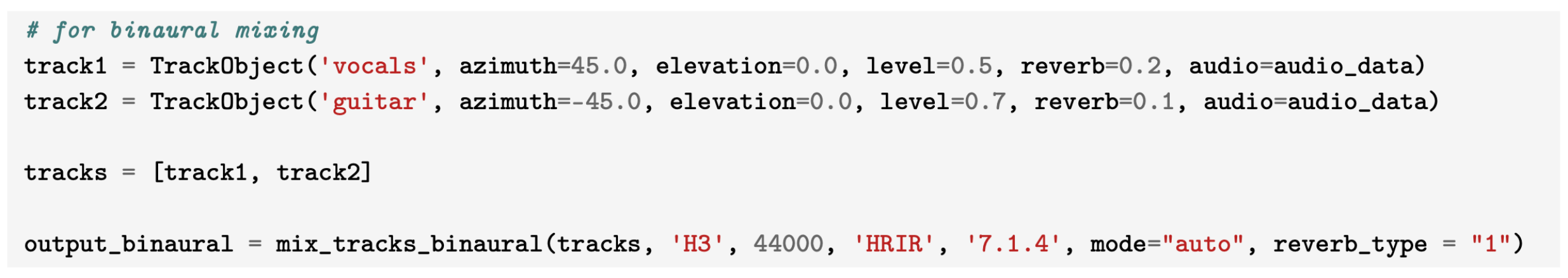}
    \caption{Example of binaural mixing}
  \label{fig:binaural_code}
\end{figure}

Figure \ref{fig:binaural_code} shows how the \texttt{TrackObject} is used and how it is passed to the binaural mixer function, \texttt{mix\_tracks\_binaural}. This function takes an array of track objects, the subject ID, sample rate, IR Type, speaker layout, interpolation mode and reverb type as arguments. There is also a separate stereo mixer function which can be used in a very similar manner but uses constant power panning instead of azimuth and elevation. Consult the Binamix documentation for more information.

Referring to the function arguments depicted in figure \ref{fig:binaural_code}, subject IDs and IR types are discussed in section \ref{sec:binamix} and full details can be found in \cite{armstrong2018hrtfs}. There are 4 reverb models to choose from: 1 = Theatre, 2 = Office, 3 = Small Room, 4 = Meeting Room. These IRs were taken from the Achen Impulse Response Database \cite{AIRdatabase}. The surround speaker layouts are discussed further in section \ref{subsec:surround}.

\subsection{Surround Rendering and Layout Simulation}
\label{subsec:surround}
\begin{figure}[ht]
  \centering
    \includegraphics[width=1\textwidth]{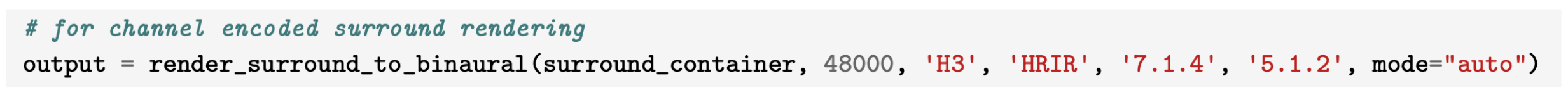}
    \caption{Example of channel-encoded surround rendering}
  \label{fig:surround_code}
\end{figure}

The \texttt{render\_surround\_to\_binaural} function works in a similar way to the \texttt{mix\_tracks\_binaural} function except that it uses an array of channel-encoded surround tracks as the input. Figure \ref{fig:surround_code} shows how the function is used.The function takes the input and output layouts as arguments (e.g. input = 7.1.4, output = 5.1.4) and binaurally renders the audio accordingly to the correct angles for the chosen layout. The angles for each layout are defined in the \texttt{surround\_utilities.py} file which currently supports the following layouts: \texttt{['5.1', '5.1.4', '5.1.2', '7.1', '7.1.4', '7.1.2', '9.1.4', '9.1.2', '9.1']}. If the input and output layout are the same, the renderer will use discrete angles for the specified surround layout. If the output format differs, the rendering function will use interpolation methods described in section \ref{subsec:interpolation} to approximate a downmix. The channel order and naming conventions come from the SMPTE ST 2067-8 standard \cite{SMPTE_ST_2067-8} and Dolby "Additional Channels For Immersive Audio" \cite{Dolby2018AdditionalChannels}.

\subsection{Interpolation Method}
\label{subsec:interpolation}
When using surround layout simulation with the binaural mixer, the tools will use IR interpolation to simulate the VBAP method \cite{Pulkki1997:VBAPbase} and render the source at the virtual speaker positions between the discrete angles available for a speaker layout. This allows you to create simulated surround mixes (binaurally rendered) for any given speaker layout described in \ref{subsec:surround}. This is not to be confused with the dedicated channel-encoded surround rendering function described in \ref{subsec:surround} which does not use interpolation, unless downmixing to a lower order speaker layout. All of this logic is handled automatically but it is important to understand the distinction. To clarify, 

\begin{itemize}
  \item \texttt{mix\_tracks\_binaural} allows you to create your own binaural mixes which can simulate the limitations of channel-encoded surround (e.g., amplitude-only panning).
  \item \texttt{render\_surround\_to\_binaural} allows you to binaurally render an existing real-world channel-encoded surround mix.
\end{itemize}

Interpolation may also be required even when speaker layout simulation is not being used. For example, there are only 50 BRIR angles available for the human subjects within SADIE II compared to the 8802 angles available for dummy subject HRIRs. So in the instance where a SADIE subject does not contain an angle within 2 degrees of the desired angle, a suitable interpolated angle is generated using proximal angles. 

\subsubsection{Delaunay Triangulation}
\begin{figure}[ht]
  \centering
  \begin{minipage}[t]{0.48\textwidth}
    \centering
    \includegraphics[width=\textwidth]{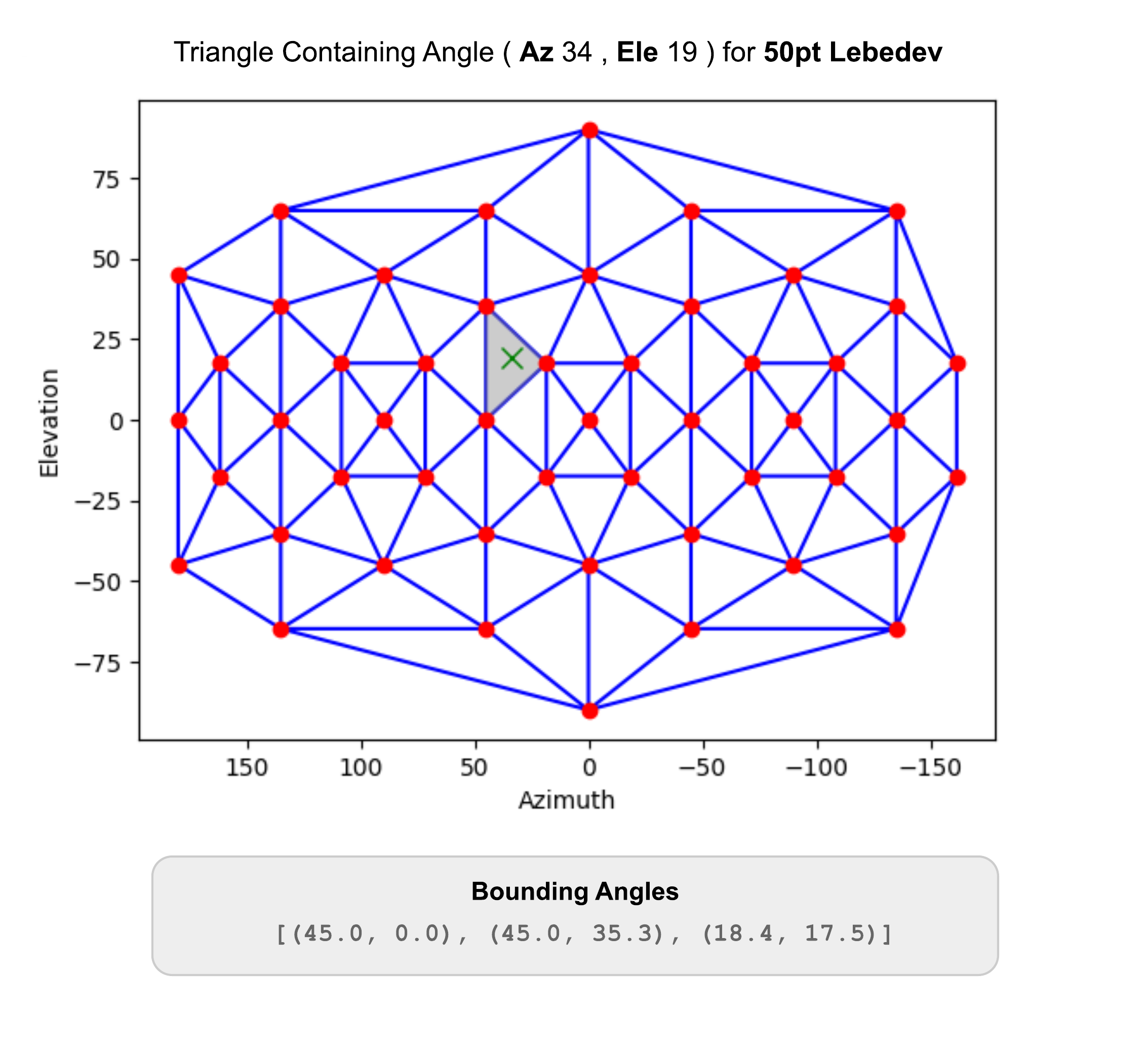}
  \end{minipage}%
  \hfill
  \begin{minipage}[t]{0.48\textwidth}
    \centering
    \includegraphics[width=\textwidth]{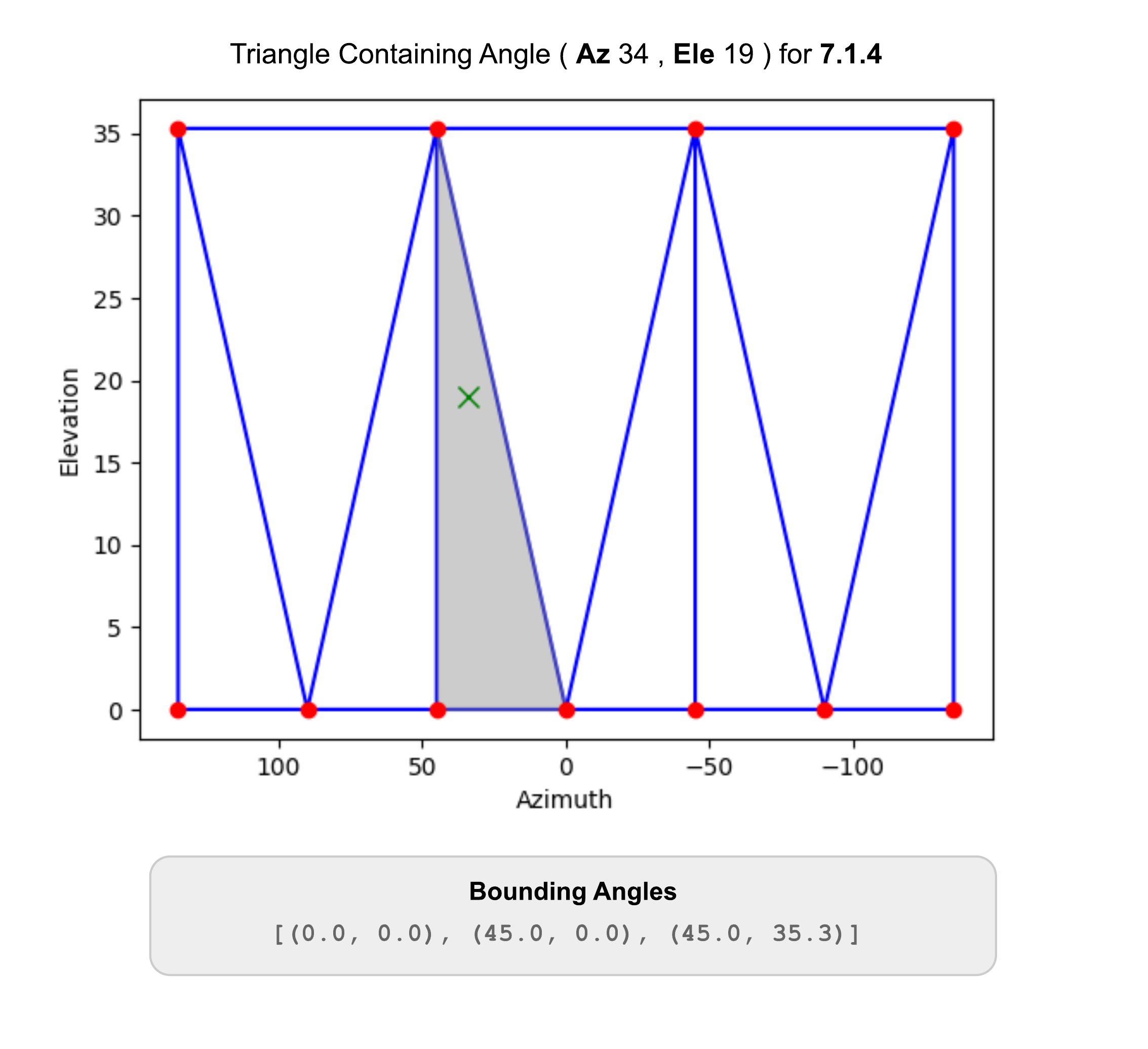}
  \end{minipage}
  \caption{Delaunay Triangulation on two spatial distributions: 50-point Lebedev (left) and 7.1.4 layout (right).}
  \label{fig:triangulations}
\end{figure}

Finding proximal angles to use for interpolation is slightly more challenging than finding the three nearest neighbors because the sphere is not uniformly sampled across azimuth and elevation. For human subject BRIRs in SADIE II, a 50 point Lebedev Quadrature \cite{Lebedev} distribution is used, and for HRIRs, higher order sampling is used but azimuth resolution is always significantly higher than elevation resolution. As result, a nearest neighbor search will almost always return 3 points on a line instead of the triangle within which the point lies. To overcome this problem, we used a modified Delaunay triangulation. The Delaunay method is normally used for 2D and 3D triangulation with cartesian coordinates but our modified version considers the angles and elevations to be an equirectangular projection of the sphere which is valid to find the bounding triangle for the desired angle. However, it is not a valid domain to calculate distances, so we convert the bounding angles into the cartesian domain. Then we calculate the Euclidean distance from the desired angle to the bounding angles in order to calculate interpolation weights for generating the required IR. The spherical domain is periodic but an equirectangular projection is not, so it is possible for a desired angle to fall outside the Delaunay mesh. If this happens, our method will rotate the projection by 180 degrees, vertically and horizontally if necessary and will always return a bounding triangle. Figure \ref{fig:triangulations} shows the Delaunay triangulation of a desired angle within a 50pt Lebedev distribution and a 7.1.4 distribution. The \texttt{delaunay\_triangulation} function in the Binamix library generates these plots by setting \texttt{plots=True} in the arguments. 

\subsubsection{Interpolation Modes}
\begin{figure}
  \centering
  \includegraphics[width=0.8\textwidth]{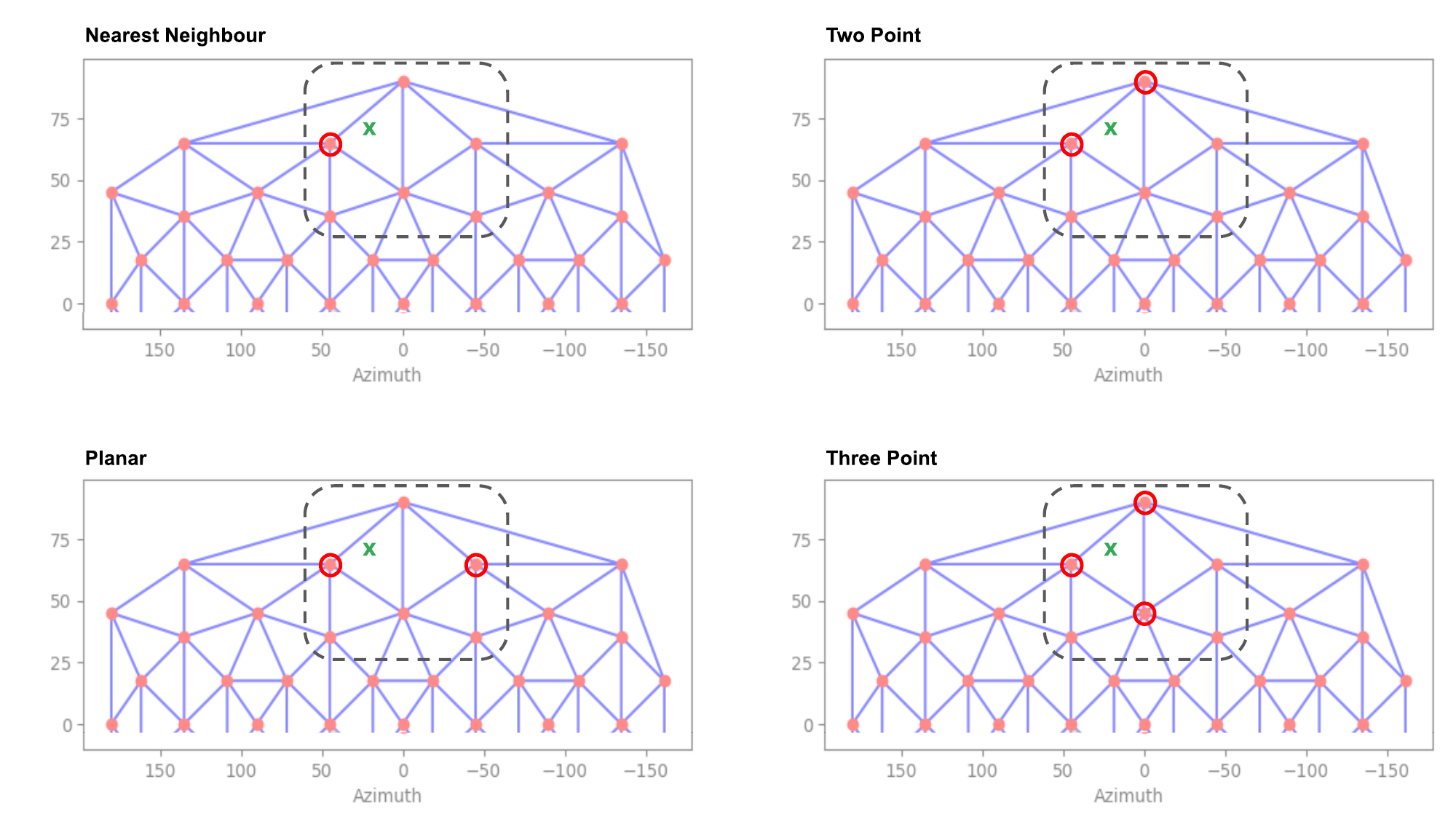}
  \caption{Interpolation Modes: Nearest Neighbor, Two Point, Three Point and Planar}
  \label{fig:interpolation}
\end{figure}

The library provides 5 interpolation modes, each of which has a slightly different effect on the resulting render. Figure \ref{fig:interpolation} depicts which points are used for each of the modes which are described as follows:

\begin{itemize}
  \item \textbf{\texttt{nearest}}: Uses the nearest available discrete angle, thus avoiding the use of interpolation at all. This mode guarantees the use of a real IR but compromises spatial fidelity for low density point distributions.
  \item \textbf{\texttt{planar}}: Creates a weighted interpolation between the nearest neighbors on the closest elevation plane. This compromises elevation accuracy but suffers less spectral coloration than three point interpolation.
  \item \textbf{\texttt{two\_point}}: Creates a weighted interpolation between the two points on the nearest line to the desired angle. The function automatically chooses between azimuth or elevation interpolation based on the point distribution or speaker layout.
  \item \textbf{\texttt{three\_point}}: Uses three-point weighted interpolation between the bounding angles found using the Delaunay triangulation method described above.
  \item \textbf{\texttt{auto}}: Automatically selects the best interpolation method based on which one achieves the closest Euclidean approximation to the desired angle.
\end{itemize}

In general, using the fewest points possible for interpolation leads to the most favorable outcome since each point contributes its own ITD, ILD and spectral cues to the interpolated IR. Furthermore, interpolation fidelity is inherently limited by the available point density.

\subsection{Example Scripts}
\label{subsec:examples}
The library also includes several example scripts which briefly demonstrate how to use the library for various tasks. Our hope is that they serve as useful templates for your own projects. Example scripts include:
\begin{itemize}
  \item \texttt{binaural\_mixer\_example.py} – Basic example of binaural mixing with 4 tracks.
  \item \texttt{render\_surround\_encoded\_wav\_example.py} – Example of rendering a multichannel surround mix to binaural.
  \item \texttt{stereo\_mixer\_example.py} – Basic example of stereo mixing with 4 tracks.
  \item \texttt{mix\_for\_all\_speaker\_layouts\_example.py} – Render a single mix for all available speaker layouts.
  \item \texttt{opus\_transcodes\_example.py} – Transcode audio files to Opus format.
  \item \texttt{triangulation\_with\_plot\_example.py} – Example of Delaunay triangulation with plot.
  \item \texttt{generate\_binaural\_dataset\_example.py} – Example of generating a binaural dataset.
\end{itemize}

\section{Conclusion}
Binamix provides a modular and extensible framework for generating high-quality binaural audio datasets at scale. By integrating robust binaural rendering, flexible mixing pipelines, and support for multiple speaker layouts, the library simplifies the process of simulating realistic binaural audio conditions. The inclusion of interpolation methods and layout-aware rendering ensures accuracy even when discrete IRs are unavailable, and the modular design allows for easy customization and integration into larger experimental workflows. 

The library is not designed to be a fully functional spatial audio production suite, but there are still some features we plan to add in the future. Currently, the library only supports mono reverb as a process prior to HRIR/BRIR convolution. This is suitable for our current needs but a multichannel reverb would allow for a more flexible simulation of real-world production conditions. The library currently only supports static source positions but we plan to add support for dynamic source movement in the future. 

We hope this tool accelerates research in spatial audio coding, quality assessment, and machine learning by lowering the barrier to reproducible and customizable binaural dataset creation.

\section{Acknowledgments}
This work was conducted with a research grant from Taighde Éireann – Research Ireland co-funded under the European Regional Development Fund under Grant Numbers 12/RC/2289\_P2 and a gift from Google. For the purpose of Open Access, the author has applied a CC BY public copyright license to any Author Accepted Manuscript version arising from this submission.

\printbibliography

\end{document}